\documentclass[12pt]{article}
\usepackage{amssymb}
\usepackage{graphicx}
\usepackage{subfigure}
\usepackage{psfrag}

\newcommand{\be}{\begin{equation}}
\newcommand{\ee}{\end{equation}}
\newcommand{\bea}{\begin{eqnarray}}
\newcommand{\eea}{\end{eqnarray}}
\newcommand{\ba}{\begin{array}}
\newcommand{\ea}{\end{array}}
\newcommand{\bi}{\begin{itemize}}
\newcommand{\ei}{\end{itemize}}
\newcommand{\bn}{\begin{enumerate}}
\newcommand{\en}{\end{enumerate}}
\newcommand{\bc}{\begin{center}}
\newcommand{\ec}{\end{center}}
\newcommand{\al}{\alpha}
\newcommand{\de}{\delta}
\newcommand{\la}{\lambda}
\newcommand{\te}{\theta}
\renewcommand{\l}{\left}
\renewcommand{\r}{\right}

\newcommand{\ol}{\overline}

\newcommand{\nl}{\nonumber\\}

\newcommand{\wt}[1]{\widetilde{#1}}

\begin{document}
\tolerance=100000


\vspace*{\fill}

\begin{center}
{\Large \bf
$B_s - \overline{B}_s$ mixing in the MSSM scenario with
large flavor mixing in the LL/RR sector
}\\[3.cm]

{\large\bf Seungwon Baek\footnote{sbaek@cskim.yonsei.ac.kr}}
\\[7mm]

{\it
Department of Physics, Yonsei University, Seoul 120-749, Korea
}\\[10mm]
\end{center}

\vspace*{\fill}

\begin{abstract}
{\small\noindent
We show that the recent measurements of $B_s-\overline{B}_s$ mass difference,
$\Delta m_s$,
by D{\O}  and CDF collaborations give very strong constraints on
MSSM scenario with large flavor mixing in the LL and/or RR sector of down-type
squark mass squared matrix. In particular, the region with
large mixing angle and large mass
difference between scalar strange and scalar bottom is ruled out
by giving too large $\Delta m_s$. The allowed region is sensitive to the
CP violating phases $\delta_{L(R)}$. The $\Delta m_s$ constraint is
most stringent on the scenario with both LL and RR mixing.
We also predict the time-dependent CP asymmetry in $B_s \to \psi \phi$ decay
and semileptonic asymmetry in $B_s\to \ell X$ decay.
}
\end{abstract}

\vspace*{\fill}


\newpage

\section{Introduction}
\label{sec:intro}

The flavor changing processes in the ${s}-{b}$ sector are sensitive probe
of new physics (NP) beyond the standard model (SM) because they are experimentally the
least constrained.
In the minimal supersymmetric standard model (MSSM), however, the
flavor mixing in the chirality flipping
down-type squarks, $\wt{s}_{L(R)}-\wt{b}_{R(L)}$, is already strongly
constrained by the measurement of $BR(B \to X_s \gamma)$. On the other hand,
large flavor mixing in the chirality conserving
$\wt{s}_{L(R)} - \wt{b}_{L(R)}$ has been largely allowed.
Especially the large mixing scenario in the $\wt{s}_R - \wt{b}_R$ sector
has been drawing much interest
because it is well motivated by the measurement large neutrino mixing
and the idea of grand unification~\cite{Baek:GUT}.

Recently D{\O} and CDF collaborations at Fermilab Tevatron reported the results
on the measurements of $B_s - \ol{B}_s$ mass difference~\cite{D0,CDF}
\bea
17 ~{\rm ps}^{-1} < \Delta m_s < 21 ~{\rm ps}^{-1} ~~(90 \% \mbox{~CL}), \nl
\Delta m_s = 17.33^{+0.42}_{-0.21}\pm 0.07 ~{\rm ps}^{-1},
\label{dms:exp}
\eea
respectively.
These measured values are consistent with the SM predictions~\cite{Bona:2005eu,CKMfitter}
\bea
\Delta m_s^{\rm SM}({\rm UTfit}) = 21.5 \pm 2.6 ~{\rm ps}^{-1}, \quad
\Delta m_s^{\rm SM}({\rm CKMfit}) = 21.7^{+5.9}_{-4.2} ~{\rm ps}^{-1}
\label{dms:SM}
\eea
which are obtained from global fits,
although the experimental measurements in (\ref{dms:exp}) are slightly lower.
The implications of $\Delta m_s$ measurements have already been considered
in model independent approach~\cite{model_indep1,model_indep2,Buras:2006}, MSSM
models~\cite{MSSM,Endo:2006dm},
$Z'$-models~\cite{Zprime}, {\it etc}.

In this paper, we consider the implications of (\ref{dms:exp}) on an MSSM
scenario with large mixing in the LL and/or RR sector. We do not consider
flavor mixing in the LR(RL) sector because they are i) are already strongly
constrained by $BR(B \to X_s  \gamma)$~\cite{LR} and ii) therefore relatively insensitive to
$B_s - \ol{B}_s$ mixing. We neglect mixing between the 1st and
2nd generations which are tightly constrained by $K$ meson decays and
$K - \ol{K}$ mixing, and mixing between the 1st and 3rd generations
which is also known to be small by the measurement of $B_d - \ol{B}_d$ mixing.

The paper is organized as follows. In Section~\ref{sec:BsBs}, the relevant
formulas for $B_s-\ol{B}_s$ mixing are presented. In Section~\ref{sec:num}
we perform numerical analysis and show the constraints imposed
on our scenario.
With these constraints, in Section~\ref{sec:asym}, we predict the
time-dependent CP asymmetry in $B_s \to \psi \phi$ decay
and the semileptonic asymmetry in $B_s\to \ell X$ decay.
We conclude in Section~\ref{sec:con}.

\section{$B_s-\ol{B}_s$ mixing in the MSSM scenario with large LL/RR mixing}
\label{sec:BsBs}

According to the description of our model in Section~\ref{sec:intro},
the scalar down-type mass squared matrix in the basis where down quark mass matrix
is diagonal is given by~\cite{GNK,Baek:Bs2KK}
\bea
 M^2_{\wt{d},LL}
=\l(\ba{ccc}
 \wt{m}^{d,2}_{L_{11}} & 0 & 0 \\
 0 & \wt{m}^{d,2}_{L_{22}} & \wt{m}^{d,2}_{L_{23}} \\
 0 & \wt{m}^{d,2}_{L_{32}} & \wt{m}^{d,2}_{L_{33}} \\
\ea
\r), \quad
 M^2_{\wt{d},LR(RL)} \equiv 0_{3\times 3}.
\eea
The  $M^2_{\wt{d},RR}$ can be obtained from $M^2_{\wt{d},LL}$
by exchanging $L \leftrightarrow R$.
We note that this kind of scenario is orthogonal to the one with flavor
violation controlled only by CKM matrix
(minimal flavor violation model~\cite{MFV,Buras:2006} or the effective
SUSY model considered in~\cite{Baek:1998yn}),
where large flavor violation in $s-b$ is impossible a priori.

The mass matrix $M^2_{\wt{d},LL}$ can be diagonalized by
\bea
 \Gamma_L M^2_{\wt{d},LL} \Gamma_L^\dagger
= {\rm diag}(m^2_{\wt{d}_L},m^2_{\wt{s}_L},m^2_{\wt{b}_L}),
\eea
with
\bea
 \Gamma_L
=\l(\ba{ccc}
 1 & 0 & 0 \\
 0 & \cos\te_L & \sin\te_L\;e^{i\de_L} \\
 0 & -\sin\te_L\;e^{-i\de_L} & \cos\te_L  \\
\ea
\r).
\label{eq:Gamma_L}
\eea
Similarly, the exchange  $L \leftrightarrow R$ in (\ref{eq:Gamma_L}) gives $\Gamma_R$.
We restrict $ -45^\circ < \te_{L(R)} < 45^\circ$ so that the
mass eigenstate ${\wt{s}(\wt{b})}$ has more strange (beauty) flavor than beauty (strange)
flavor.

The most general effective Hamiltonian for $B_s - \ol{B}_s$ mixing
\bea
  H_{\rm eff} = \sum_{i=1}^5 C_i O_i + \sum_{i=1}^3 \wt{C}_i \wt{O}_i
\eea
has 8 independent operators
\bea
 O_1 &=& (\ol{s}_L \gamma_\mu b_L) \;(\ol{s}_L \gamma^\mu b_L), \nl
 O_2 &=& (\ol{s}_R b_L) \;(\ol{s}_R b_L), \nl
 O_3 &=& (\ol{s}_R^\alpha b_L^\beta) \;(\ol{s}_R^\beta b_L^\alpha), \nl
 O_4 &=& (\ol{s}_R b_L) \;(\ol{s}_L b_R), \nl
 O_5 &=& (\ol{s}_R^\alpha b_L^\beta) \;(\ol{s}_L^\beta b_R^\alpha), \nl
 \wt{O}_{i=1,\cdots 3} &=&  O_{i=1,\cdots 3} \l|_{L \leftrightarrow R}\r..
\eea

The Wilson coefficients for these
$\Delta B = \Delta S =2$ operators can be obtained
by calculating the gluino mediated box diagrams. Since the chargino and neutralino
exchanged box diagrams are suppressed by the small gauge coupling constants,
we neglect them. In the scenario we are considering, when we consider only LL (RR)
mixing, the SUSY box diagram contributes only to $C_1$ ($\wt{C}_1$).
When both LL and RR mixing exist simultaneously, there are also contributions
to $C_4$ and $C_5$. However, ${\stackrel{(\sim)}{C}}_2$ or
${\stackrel{(\sim)}{C}}_3$ are not generated at all.
Note that the induced LR (RL) mixing~\cite{Baek:1999} does not occur, either,
because we set $M^2_{\wt{d},LR(RL)} \equiv 0_{3\times 3}$. Otherwise,
the SUSY parameter space is further constrained depending
on $\tan\beta$~\cite{Baek:1999}.
The analytic formulas for the Wilson coefficients at the MSSM scale
are given by
\bea
 C_1^{\rm MSSM} &=& {\alpha_s^2 \over 4 m^2_{\wt{g}}} \sin^2 2\te_L e^{2i \de_L}
     \l(
f_1(x_{\wt{b}_L,\wt{g}},x_{\wt{b}_L,\wt{g}})
-2 f_1(x_{\wt{s}_L,\wt{g}},x_{\wt{b}_L,\wt{g}})
+f_1(x_{\wt{s}_L,\wt{g}},x_{\wt{s}_L,\wt{g}})
 \r),
\nl
 C_{4(5)}^{\rm MSSM} &=& {\alpha_s^2 \over 4 m^2_{\wt{g}}} \sin 2\te_L \sin 2\te_R
 e^{i (\de_L+\de_R)}
     \l(
f_{4(5)}(x_{\wt{b}_R,\wt{g}},x_{\wt{b}_L,\wt{g}})
-f_{4(5)}(x_{\wt{b}_R,\wt{g}},x_{\wt{s}_L,\wt{g}}) \r.\nl
&&\l.-f_{4(5)}(x_{\wt{s}_R,\wt{g}},x_{\wt{b}_L,\wt{g}})
+f_{4(5)}(x_{\wt{s}_R,\wt{g}},x_{\wt{s}_L,\wt{g}})
 \r), \nl
 \wt{C}_1^{\rm MSSM} &=&  C_1^{\rm MSSM} \l|_{L \leftrightarrow R}\r.,
\label{eq:WC}
\eea
where the loop functions are defined as
\bea
 f_1(x,y) &\equiv& {1 \over 9} j(1,x,y) + {11 \over 36} k(1,x,y), \nl
 f_4(x,y) &\equiv& {7 \over 3} j(1,x,y) - {1 \over 3} k(1,x,y), \nl
 f_5(x,y) &\equiv& {1 \over 9} j(1,x,y) + {5 \over 9} k(1,x,y),
\eea
and the $j$ and $k$ are defined in~\cite{Colangelo:1998pm}.
The RG running of the Wilson coefficients down to $m_b$ scale can be found,
for example, in~\cite{Becirevic:2001jj}.

We can calculate the $B_s - \ol{B}_s$ mixing matrix element, which is
in the form
\bea
 M^s_{12} = M_{12}^{s,\rm SM} (1 + R).
\label{eq:M12}
\eea
The mass difference of $B_s-\ol{B}_s$ system is then given by
\bea
\Delta m_s &=& 2 | M^s_{12} | \nl
           &=& \Delta m_s^{\rm SM} | 1 + R |.
\label{eq:dms}
\eea
In the SM contribution~\cite{Buras:1990fn} to the mass matrix element
\bea
 M_{12}^{s,\rm SM} = {G_F^2 M_W^2 \over 12 \pi^2}
  M_{B_s} \l(f_{B_s} \hat{B}_{B_s}^{1/2} \r)^2
  \eta_B S_0(x_t) \l(V_{tb} V_{ts}^*\r)^2,
\label{eq:M12_SM}
\eea
the non-perturbative parameters $f_{B_s}$ and $\hat{B}_{B_s}$
give main contribution to the theoretical uncertainty.
Using the combined lattice result~\cite{Okamoto:2005zg}
from JLQCD~\cite{Aoki:2003xb} and HPQCD~\cite{Gray:2005ad},
\bea
  f_{B_s} \hat{B}_{B_s}^{1/2} \Bigg|_{\rm (HP+JL)QCD} = (0.295 \pm 0.036) \;\; {\rm GeV},
\eea
the SM predicts
\bea
 \Delta m_s^{\rm SM} = (22.5 \pm 5.5) \;\; {\rm ps}^{-1},
\label{eq:dms_SM_pre}
\eea
which is consistent with the values in (\ref{dms:SM}) obtained from global
fits.
For the prediction in (\ref{eq:dms_SM_pre}), we used $\eta_B = 0.551$,
$\ol{m}^{\ol{MS}}_t(m_t) = 162.3$ GeV and
$V_{ts}=0.04113$~\cite{Charles:2006yw}.

Now, inserting the CDF data in (\ref{dms:exp}) and the
SM prediction in (\ref{eq:dms_SM_pre}) into (\ref{eq:dms}), we obtain
\bea
 |1+R| = 0.77 ^{+0.02}_{-0.01}({\rm exp}) \pm 0.19 ({\rm th}),
\label{eq:R}
\eea
where the experimental and theoretical errors were explicitly written.
The expression for $R$ in our scenario is given
by~\footnote{The $\hat{B}_{B_s}$ in (\ref{eq:M12_SM}) is related to
$B_1(\mu_b)$ as~\cite{Buras:1990fn}
\bea
\hat{B}_{B_s} &\equiv& B_1(\mu_b) [\al_s^{(5)}(\mu_b)]^{-6/23}
 \l[1+{\al_s^{(5)}(\mu_b) \over 4 \pi} J_5\r].
\eea
}
\bea
 R(\mu_b) &=& \xi_1(\mu_b) + \wt{\xi}_1(\mu_b) +
  {3 \over 4} {B_4(\mu_b) \over B_1(\mu_b)}\l(M_{B_s} \over m_b(\mu_b) + m_s(\mu_b)\r)^2
 \xi_4 \nl
 && + {1 \over 4}{B_5(\mu_b) \over B_1(\mu_b)}\l(M_{B_s} \over m_b(\mu_b) + m_s(\mu_b)\r)^2
\xi_5,
\eea
where we defined ($i=1,\cdots,5$)
\bea
\xi_i(\mu_b) &\equiv& C_i^{\rm SUSY}(\mu_b)/C_1^{\rm SM}(\mu_b), \nl
\wt{\xi_i}(\mu_b) &\equiv& \wt{C}_i^{\rm SUSY}(\mu_b)/C_1^{\rm SM}(\mu_b).
\eea
The relevant B-parameters are given in~\cite{Becirevic:2001xt} by
\bea
 B_1(\mu_b) = 0.86(2)\l(^{+5}_{-4}\r), \quad
 B_4(\mu_b) = 1.17(2)\l(^{+5}_{-7}\r), \quad
 B_5(\mu_b) = 1.94(3)\l(^{+23}_{-7}\r).
\eea

Now we briefly discuss $B \to X_s  \gamma$ constraint.
The SUSY parameters we consider are also directly constrained by the measured
branching ratio of inclusive radiative $B$-meson decay, $B \to X_s  \gamma$.
We take this constraint
into account, although it is not expected to be so
severe as in a scenario with LR or RL mixing.
In the operator basis given in \cite{QCD_anatomy}, the
SUSY contributions to the Wilson coefficients of magnetic operators
in our scenario are
\bea
C^{\rm SUSY}_{7\gamma} &=& -{4 \over 9} \; {1 \over \la_t}\;
  { \pi \al_s \sin 2\te_L e^{i \de_L} \over \sqrt{2} G_F m^2_{\wt{g}} }
 \l[J_1(x_{b_L g}) - J_1(x_{s_L g}) \r], \nl
C^{\rm SUSY}_{8g} &=&  \; {1 \over \la_t}\;
  { \pi \al_s \sin 2\te_L e^{i \de_L} \over \sqrt{2} G_F m^2_{\wt{g}} }
 \l[\l(-{3 \over 2} I_1(x_{b_L g}) -{1 \over 6} J_1(x_{b_L
 g})\r)
  -(b_L \leftrightarrow s_L ) \r],
\label{eq:C7C8}
\eea
where $\la_t=V_{ts}^* V_{tb}$ and
\bea
  I_1(x) &=& \frac{1-6x+3x^2+2x^3-6x^2 \log x}{12(1-x)^4}, \nl
  J_1(x) &=& \frac{2+3x-6x^2+x^3+6x\log x}{12(1-x)^4}.
\eea
There are also chirality flipped $\wt{C}_{7\gamma,8g}$ with $L$
replaced by $R$.
Therefore, we can see that in principle $\te_{L(R)}$,$\de_{L(R)}$ and
$m_{\wt{s}}-m_{\wt{b}}$ can be constrained.
Compared to the $LR(RL)$ mixing case where large SUSY contribution
${\cal O}(m_{\wt{g}}/m_b)$ is
possible due to the chirality flipping inside the loop,
our scenario allows only a small SUSY correction to the
SM contributions.
In addition, although LL mixing gives a linear correction
${\cal O}(C_{7\gamma,8g}^{\rm SUSY}/C_{7\gamma,8g}^{\rm SM})$
due to the interference term,
RR mixing generates only a quadratic correction
${\cal O}(|C_{7\gamma,8g}^{\rm SUSY}/C_{7\gamma,8g}^{\rm SM}|^2)$
because it is added incoherently to the SM contribution.

\section{Numerical analysis}
\label{sec:num}

In this Section, we perform numerical analysis and show the
constraints imposed by $\Delta m_s^{\rm exp}$.
We also consider the $BR(B \to X_s  \gamma)$ constraint.

\begin{figure}[tbh]
\begin{center}
\psfrag{mmssLL}{$m_{ \wt{s}_L} ({\rm TeV})$}
\psfrag{ttLL}{$\te_L$}
\subfigure[]{
\includegraphics[width=0.45\textwidth]{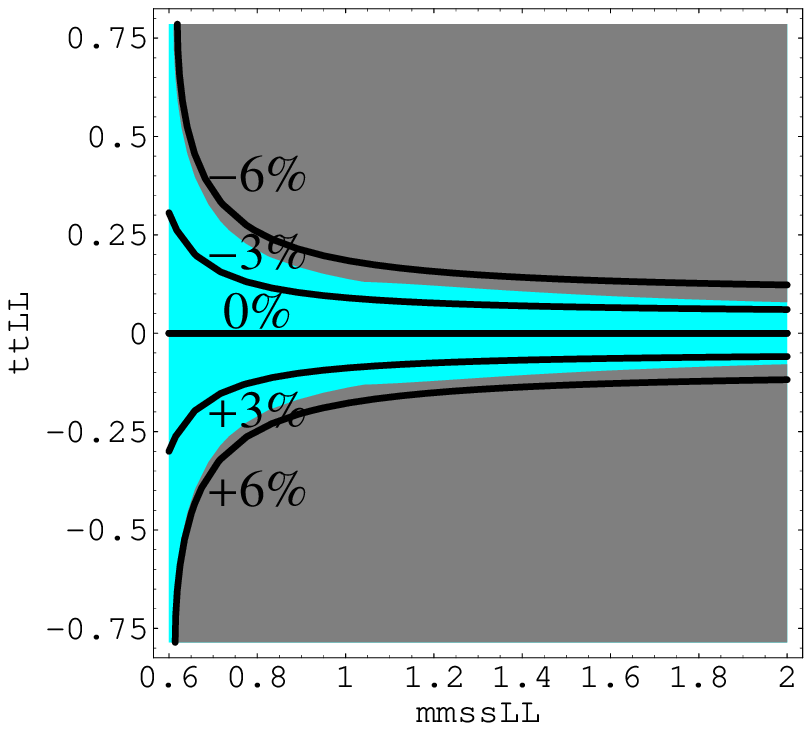}
\label{fig:teL-msL-a}
}
\subfigure[]{
\includegraphics[width=0.45\textwidth]{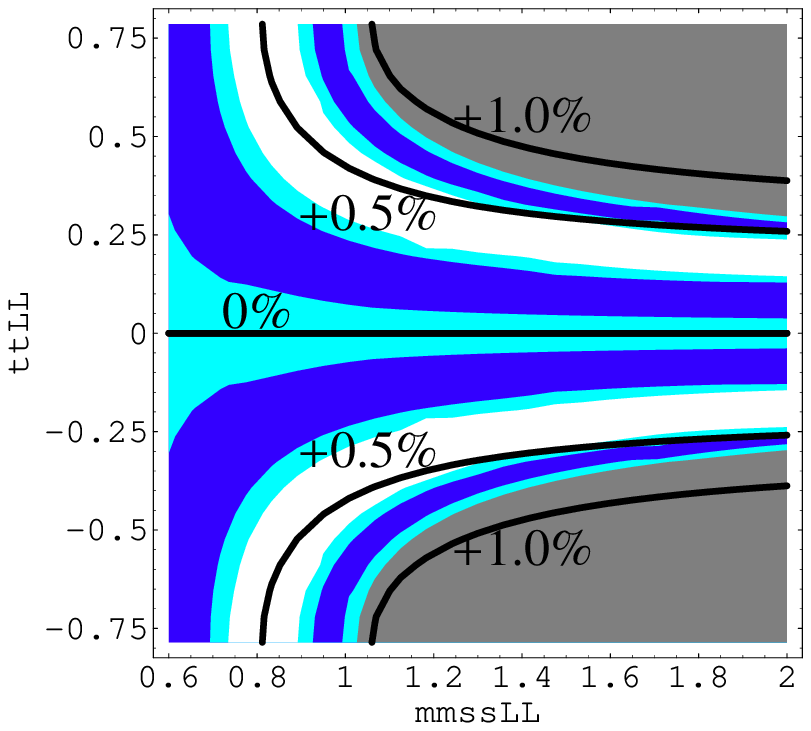}
\label{fig:teL-msL-b}
}
\end{center}
\caption{
Contour plots for $|1+R|$ in ($m_{\wt{s}_L}$,$\te_L$) plane.
Sky blue region represents
2$\sigma$ allowed region ($0.39 \le |1+R| \le 1.15$),
blue 1$\sigma$ allowed region ($0.58 \le |1+R| \le 0.96$),
and white (grey) region is excluded at 95\% CL by giving too small (large)
$\Delta m_s$.
The labeled thick lines represent
the constant
$\Big(BR^{\rm tot}(B \to X_s  \gamma)-BR^{\rm SM}(B \to X_s  \gamma)\Big)/
BR^{\rm SM}(B \to X_s  \gamma)$ contours.
Only LL mixing
is assumed to exist. The fixed parameters
are $m_{\wt{g}}=0.5$ (TeV),  $m_{\wt{b}_L}=0.5$ (TeV),
(a) $\de_L$=0, (b) $\de_L=\pi/2$.
}
\label{fig:teL-msL}
\end{figure}

From (\ref{eq:WC}) it is obvious that the larger the mass splitting between
$\wt{s}$ and $\wt{b}$, the larger the SUSY contributions are.
Therefore we expect
that (\ref{eq:R}) constrains the mass splitting when the mixing
angle $\te_{L(R)}$
is large. This can be seen in Figure~\ref{fig:teL-msL} where we show
filled contour plots for $|1+R|$ in ($m_{\wt{s}_L}$,$\te_L$) plane:
sky blue region represents
2$\sigma$ allowed region ($0.39 \le |1+R| \le 1.15$),
blue 1$\sigma$ allowed region ($0.58 \le |1+R| \le 0.96$),
and white (grey) region is excluded at 95\% CL by giving too small (large)
$\Delta m_s$.
For these plots we assumed that
only LL mixing exists and fixed $m_{\wt{g}}=0.5$ TeV, $m_{\wt{b}_L} = 0.5$ TeV.
In Figure~\ref{fig:teL-msL-a}, we fixed $\de_L = 0$. We can see that
the SUSY interferes with the SM contribution constructively
({\it i.e.} the SUSY contribution has the same sign with the SM),
and when the mixing angle
is maximal, {\it i.e.} $\te_L = \pm\pi/4$, $m_{\wt{s}_L} - m_{\wt{b}_L}$
cannot be
greater than about 150 GeV. In Figure~\ref{fig:teL-msL-b},
we set $\de_L = \pi/2$.
The SUSY contribution can interfere destructively
({\it i.e.} in opposite sign)
with the SM and much larger mass splitting
is allowed. Therefore we can see that the allowed parameters
are sensitive to the
CPV phase.

Also the constant
$\Big(BR^{\rm tot}(B \to X_s  \gamma)-BR^{\rm SM}(B \to X_s  \gamma)\Big)/
BR^{\rm SM}(B \to X_s  \gamma)$
lines are shown.
For fixed $\te_L$, larger mass splitting
$m_{\wt{s}_L}-m_{\wt{b}_L}$ gives larger deviation for the branching
ratio.
This can be understood from (\ref{eq:C7C8}).
However, for very large mass splitting the SUSY contribution
decouples and the deviation eventually
saturates.
We can see that $BR^{\rm tot}(B \to X_s  \gamma)$
deviates from the SM predictions at most about 5\% in the region
allowed by $\Delta m_s$. Since
\bea
  BR^{\rm exp}(B \to X_s  \gamma)/   BR^{\rm SM}(B \to X_s  \gamma)
=1.06 \pm 0.13
\eea
for $E_\gamma > 1.6 $ GeV~\cite{b2sr},
it is clear that the
$BR(B \to X_s  \gamma)$ constraint is completely irrelevant
in Figure~\ref{fig:teL-msL}.

The plots for the scenario with
RR mixing only are the same with Figure~\ref{fig:teL-msL} because
the expression for $B_s-\ol{B}_s$ is completely symmetric
under $L \leftrightarrow R$.
As mentioned above, the contribution to $BR(B \to X_s  \gamma)$ is
much smaller than LL case.

\begin{figure}[tbh]
\begin{center}
\psfrag{ttLL}{$\te_L$}
\psfrag{ddLL}{$\de_L$}
\subfigure[]{
\includegraphics[width=0.45\textwidth]{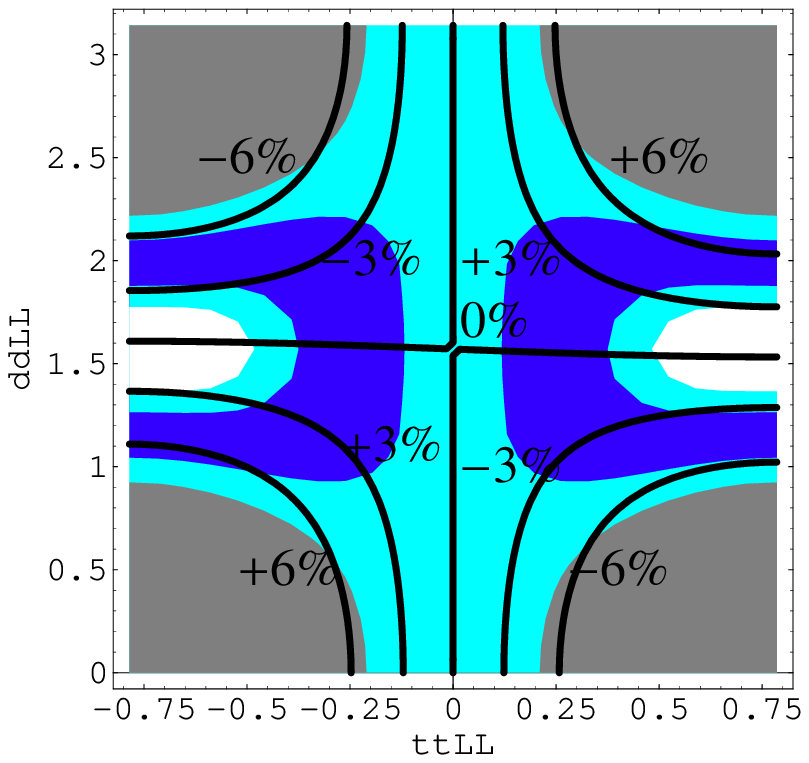}
\label{fig:teL-deL-a}
}
\subfigure[]{
\includegraphics[width=0.45\textwidth]{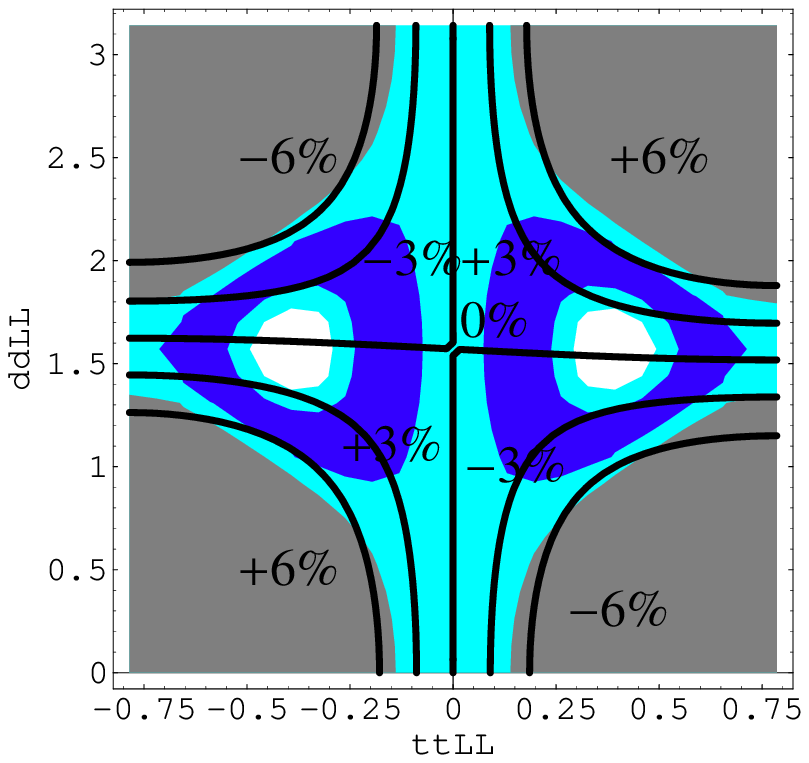}
\label{fig:teL-deL-b}
}
\end{center}
\caption{Contour plots for $|1+R|$ in ($\te_L$,$\de_L$) plane.
(a) $m_{\wt{s}_L}=0.8$ (TeV),
(b) $m_{\wt{s}_L}=1.0$ (TeV).
The rest is the same with Figure~\ref{fig:teL-msL}.
}
\label{fig:teL-deL}
\end{figure}

In Figure~\ref{fig:teL-deL}, contour plots for
constant $|1+R|$ in ($\te_L$,$\de_L$) plane
are shown. For Figure~\ref{fig:teL-deL-a}(\ref{fig:teL-deL-b}),
we fixed $m_{\wt{s}_L}=0.8(1.0)$ TeV.
The other parameters used are the same with those in Figure~\ref{fig:teL-msL}.
We can again see the strong dependence on the CPV phase $\de_L$. It can also be seen
that the parameter space with
large mixing angle $\te_L$ can be made consistent with the experiments by
cancellation with
the SM contributions in the destructive interference region
({\it i.e.} $\delta_L \approx \pi/2$).

\begin{figure}[tbh]
\begin{center}
\psfrag{ttLL}{$\te_L$}
\psfrag{ttRR}{$\te_R$}
\subfigure[]{
\includegraphics[width=0.45\textwidth]{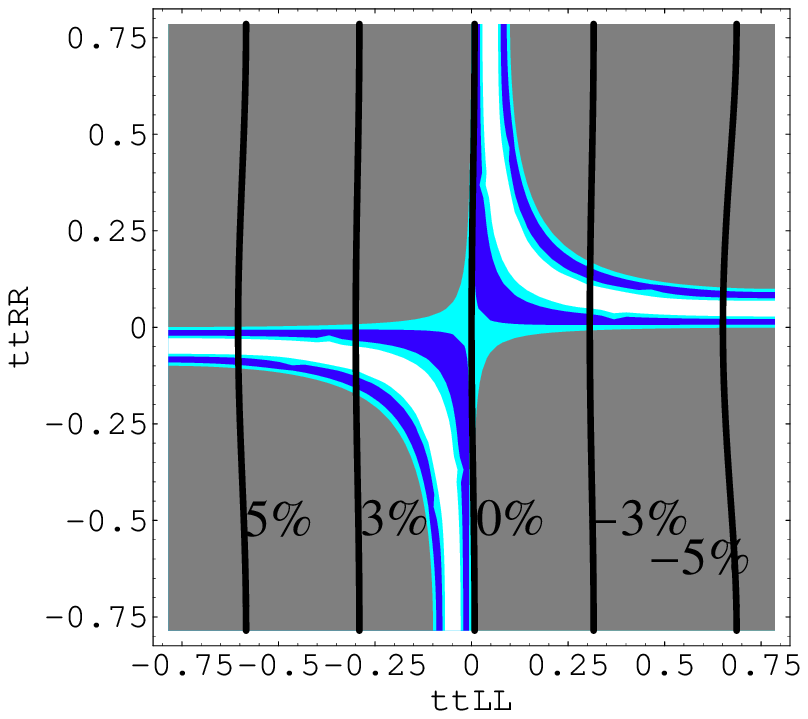}
\label{fig:teL-teR-a}
}
\subfigure[]{
\includegraphics[width=0.45\textwidth]{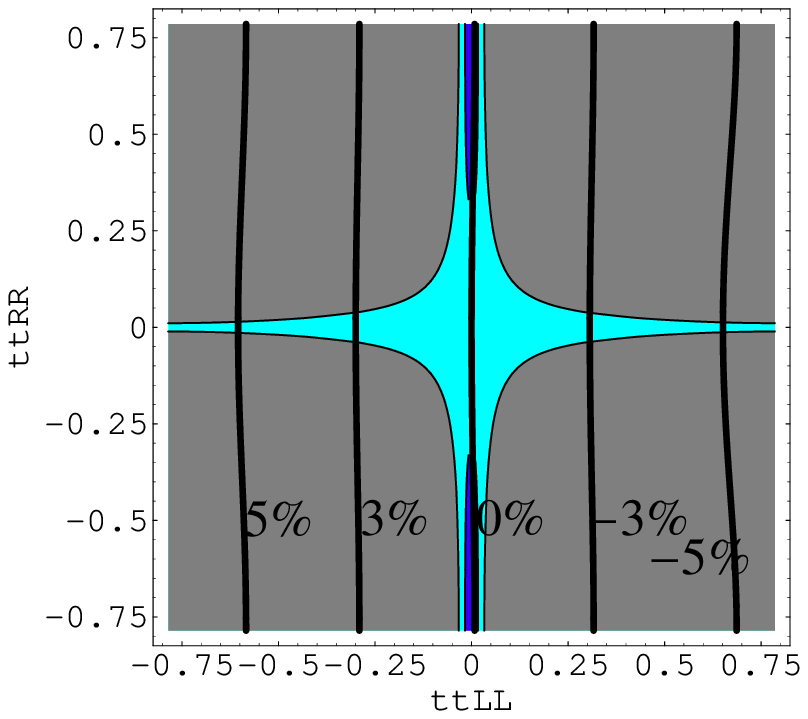}
\label{fig:teL-teR-b}
}
\end{center}
\caption{Contour plots for $|1+R|$ in ($\te_L$,$\te_R$) plane.
$m_{\wt{s}_L}=m_{\wt{s}_R}=0.6$ (TeV).
(a) $\de_L = \de_R = 0$
(b) $\de_L = 0, \de_R = \pi/2$.
We assume both LL and RR mixing exist.
The rest is the same with Figure~\ref{fig:teL-msL}.
}
\label{fig:teL-teR}
\end{figure}

Now we consider a scenario with both LL and RR mixing at the same time.
Then the operators $O_4$ and $O_5$ are additionally generated as
mentioned above.
They dominate $O_1$ or $\wt{O}_1$ for sizable mixing angles.
As a consequence, the constraint on the SUSY parameter space is very stringent as
can be seen in Figure~\ref{fig:teL-teR}.
In Figure~\ref{fig:teL-teR} we set $m_{\wt{g}}=0.5$ TeV,
$m_{\wt{b}_L}=m_{\wt{b}_R}=0.5$ TeV,
$m_{\wt{s}_L}=m_{\wt{s}_R}=0.6$ TeV, and
(a) $\de_L = \de_R = 0$ (b) $\de_L =0,  \de_R = \pi/2$.
Even for small mass splitting most region of the parameter space is
ruled out by giving too large $\Delta m_s$.
We can see that $BR(B \to X_s \gamma)$ is almost insensitive to
the change of $\theta_R$ as mentioned before.

\section{The predictions of $S_{\psi\phi}$ and $A_{\rm SL}^s$}
\label{sec:asym}

\begin{figure}[tbh]
\begin{center}
\psfrag{mmssLL}{$m_{ \wt{s}_L} ({\rm TeV})$}
\psfrag{ddLL}{$\de_L$}
\subfigure[]{
\includegraphics[width=0.45\textwidth]{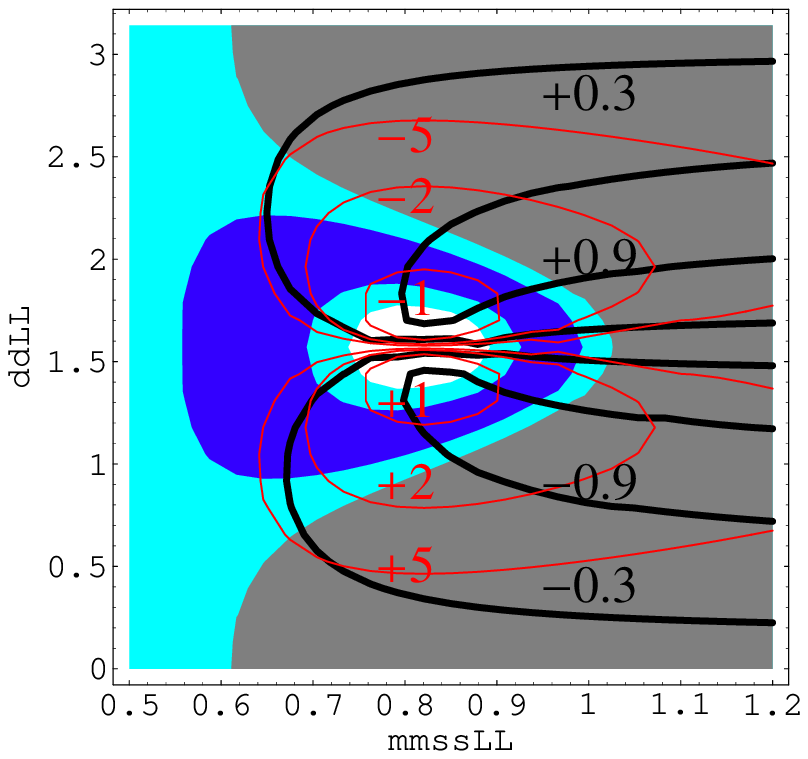}
\label{fig:msL-deL-a}
}
\subfigure[]{
\includegraphics[width=0.45\textwidth]{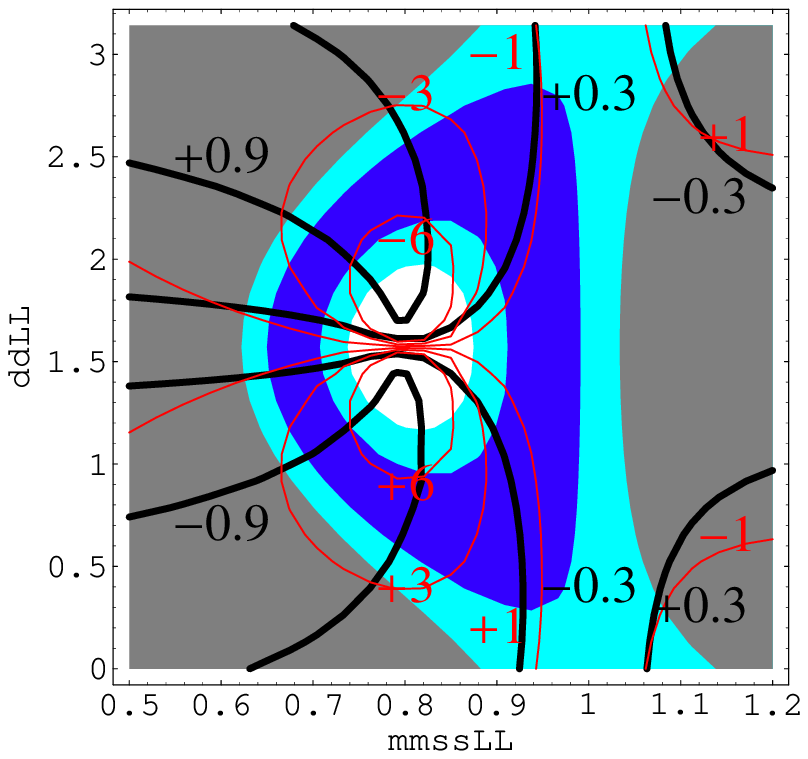}
\label{fig:msL-deL-b}
}
\end{center}
\caption{Contour plots for $|1+R|$ in ($m_{\wt{s}_L}$,$\de_L$) plane.
The $S_{\psi \phi}$ predictions are also shown as thick contour lines.
The thin red lines are constant $A_{SL}^s[10^{-3}]$ contours assuming
${\rm Re}(\Gamma_{12}^s / M^s_{12})^{\rm SM} = -0.0040$.
(a) Only LL mixing is assumed to exist.
We fixed $m_{\wt{g}}=m_{\wt{b}_L}=0.5$ TeV, $\de_L = \pi/4$.
(b) Both LL and RR mixing are assumed to exist simultaneously.
We fixed $m_{\wt{g}}=2$ TeV,
$m_{\wt{b}_L}=m_{\wt{b}_R} = 1$ TeV,
$m_{\wt{s}_R} = 1.1$ TeV,
$\te_R = \pi/4$,
$\de_L = \pi/4$, and
$\de_R = \pi/2$.
The rest is the same with Figure~\ref{fig:teL-msL}.
}
\label{fig:msL-deL}
\end{figure}

The CPV phase in the  $B_s - \ol{B}_s$ mixing amplitude will be measured
at the LHC in the near future through the time-dependent CP
asymmetry
\bea
 \frac{\Gamma(\ol{B}_s(t) \to \psi\phi)-\Gamma(B_s(t) \to \psi\phi)}
{\Gamma(\ol{B}_s(t) \to \psi\phi)+\Gamma(B_s(t) \to \psi\phi)}
\equiv S_{\psi\phi} \sin(\Delta m_s t).
\eea
In the SM, $S_{\psi\phi}$ is predicted to be very small,
$S_{\psi\phi}^{\rm SM} = -\sin 2 \beta_s = 0.038 \pm 0.003$
($\beta_s \equiv \arg[(V_{ts}^* V_{tb})/(V_{cs}^* V_{cb})]$)~\cite{model_indep2}.
If the NP has additional CPV phases, however, the prediction
\bea
 S_{\psi\phi} = -\sin(2 \beta_s + \arg(1 + R))
\eea
can be significantly different from the SM prediction.

In Figure~\ref{fig:msL-deL}, we show $|1+R|$ constraint and the prediction
of $S_{\psi\phi}$  in $(m_{\wt{s}_L},\de_L)$ plane.
However, the $B\to X_s \gamma$ prediction is not shown from now on
because it is irrelevant as mentioned above.
For Figure~\ref{fig:msL-deL-a}, we assumed the scenario with LL mixing only
and maximal mixing $\te_L = \pi/4$.
We fixed $m_{\wt{g}}=0.5$ TeV, $m_{\wt{b}_L} = 0.5$ TeV.
For Figure~\ref{fig:msL-deL-b}, we allowed both LL and RR mixing simultaneously,
while fixing
$m_{\wt{g}}=2$ TeV,
$m_{\wt{b}_L}=m_{\wt{b}_R} = 1$ TeV,
$m_{\wt{s}_R} = 1.1$ TeV,
$\te_R = \pi/4$,
$\de_L = \pi/4$, and
$\de_R = \pi/2$.
In both cases we can see that large $S_{\psi\phi}$ is
allowed for large mass splitting
between $m_{\wt{b}_L}$ and $m_{\wt{s}_L}$.
At the moment, $S_{\psi\phi}$
can take any value in the range $[-1,1]$ even after
imposing the current $\Delta m_s^{\rm exp}$ constraint.

\begin{figure}[tbh]
\begin{center}
\psfrag{SSpsiphi}{$S_{\psi\phi}$}
\psfrag{AASL}{$A_{SL}^s[10^{-3}]$}
\includegraphics[width=0.8\textwidth]{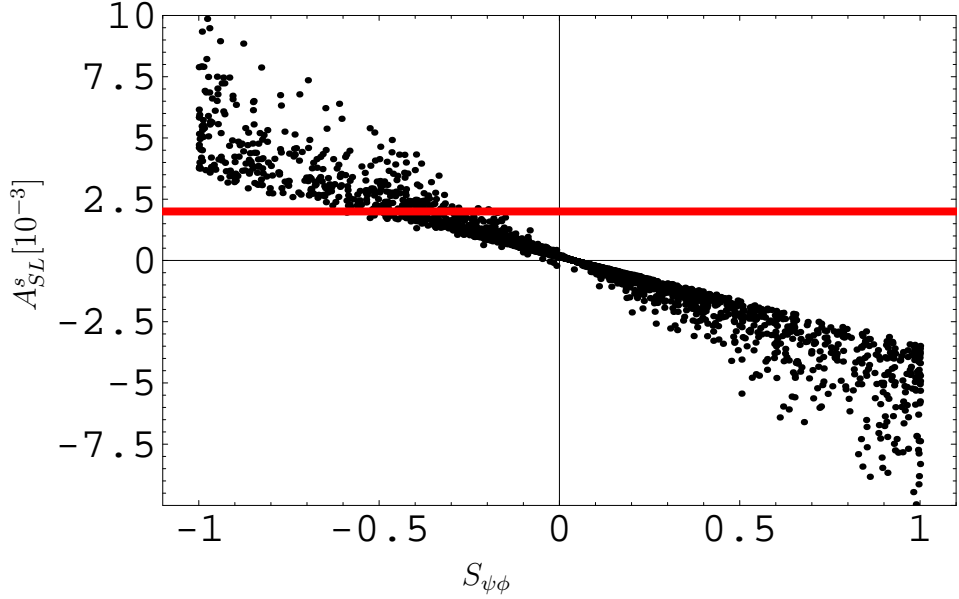}
\end{center}
\caption{
The correlation between $A^s_{\rm SL}$ and $S_{\psi\phi}$.
The red line is 1-$\sigma$ upper bound.
}
\label{fig:ASL}
\end{figure}

Finally we consider the semileptonic CP asymmetry~\cite{Randall:1998te,
Baek:1998yn,model_indep2}
\bea
 A^s_{\rm SL} \equiv
\frac{\Gamma(\ol{B}_s\to \ell^+ X)-\Gamma(B_s\to \ell^- X)}
{\Gamma(\ol{B}_s\to \ell^+ X)+\Gamma(B_s\to \ell^- X)}
={\rm Im}\l(\Gamma_{12}^s \over M^s_{12}\r).
\eea
It is approximated to be~\cite{model_indep2}
\bea
 A^s_{\rm SL} \approx {\rm Re}\l(\Gamma_{12}^s \over M^s_{12}\r)^{\rm SM}
   {\rm Im}\l(1 \over 1 + R\r),
\eea
where
${\rm Re}(\Gamma_{12}^s / M^s_{12})^{\rm SM}
= -0.0040 \pm 0.0016$~\cite{Beneke:2003}.
The SM prediction is
$A^s_{\rm SL}(\rm SM)=(2.1 \pm 0.4)\times 10^{-5}$~\cite{Beneke:2003,Ciuchini:2003}.

In Figure~\ref{fig:msL-deL}, the thin red lines are constant
$A_{SL}^s [10^{-3}]$ contours taking
${\rm Re}(\Gamma_{12}^s / M^s_{12})^{\rm SM} = -0.0040$.
We can readily see that the strong correlation between $S_{\psi\phi}$
and $A_{SL}^s$.
This can be seen from the relation
\bea
  A_{SL}^s =
- \l|{\rm Re}\l(\Gamma^s_{12} \over M^s_{12}\r)^{\rm SM}\r|
{S_{\psi\phi} \over |1+R|}.
\eea
For small $R$ the two observables are linearly correlated
as can be seen in Figure~\ref{fig:msL-deL}.

In Figure~\ref{fig:ASL}, we show the correlation
between $A^s_{\rm SL}$ and $S_{\psi\phi}$.
 We
scanned $0.5 \le m_{\wt{g}} \le 4.0$ TeV,
$0.5 < m_{\wt{b}_{L}}, m_{\wt{s}_{L}} < 2.0$ TeV,
 $-\pi/4 < \te_L < \pi/4$
and $0 < \de_L < 2 \pi$, while fixing $m_{\wt{g}}=m_{\wt{b}_L}=0.5$ TeV.
The $\Delta m_s$ constraint is imposed with $0.39 \le |1+R| \le 1.15$.
We have checked that in the scenario with only LL (RR) mixing,
we get the similar correlations.
The red line is experimental 1-$\sigma$ upper bound from
$A^s_{\rm SL} = -0.013 \pm 0.015$~\cite{model_indep2}.
Now several comments are in order:
i) The values for $S_{\psi\phi}$ and $A^s_{\rm SL}$
can be significantly different from the SM predictions.
ii)
The two observables are strongly correlated.
These two facts were already noted in \cite{model_indep2}.
It has been checked that in the $({\rm Re} R, {\rm Im} R)$ plane
the above scanned points can completely fill the
region allowed by $\Delta m_s$.
This explains why the correlation in Figure~\ref{fig:ASL} is
basically the same with model-independent prediction
in \cite{model_indep2}.
 iii)
Although it looks like that large negative $S_{\psi\phi}$ value is
disfavored, due to large error in
${\rm Re}(\Gamma_{12}^s / M^s_{12})^{\rm SM}$
we cannot definitely rule out the region at the moment.

\section{Conclusions}
\label{sec:con}

We considered the MSSM scenario with large LL and/or RR mixing
in the down-type mass squared matrix. This scenario
is strongly constrained by the recent mesurements of $B_s - \ol{B}_s$
mass difference, $\Delta m_s$, in contrast with the MSSM
scenario where the flavor mixing is controlled only by the CKM
matrix~\cite{Baek:1998yn,Buras:2006}.
The constraint is most stringent when both LL and RR mixing exist
simultaneously. It is also shown that the allowed region is quite
sensitive to the CP violating
phase.

We also considered the time-dependent CP asymmetry, $S_{\psi\phi}$, and
the semileptonic CP asymmetry, $A_{\rm SL}^s$. It was shown that
the $S_{\psi\phi}$ and $A^s_{\rm SL}$ can take values significantly
different from the SM predictions. There is also strong
correlation between $S_{\psi\phi}$ and $A_{\rm SL}^s$.

\vskip1.3cm
\noindent {\bf Acknowledgment}\\
The author thanks P.~Ko for useful discussions and the organizers
of ``Workshop on Physics at Hadron Colliders'' at KIAS where part
of this work was motivated. The work was supported by the
Korea Research Foundation Grant funded by the Korean Government
(MOEHRD) No. KRF-2005-070-C00030.

\end{document}